\documentclass[10pt,twocolumn,twoside]{IEEEtran}
\pdfoutput=1 
\usepackage{graphicx}
\usepackage[english]{babel}
\usepackage{amssymb}
\usepackage{pmat}
\usepackage{balance}
\usepackage{subfigure}
\usepackage{multirow}
\usepackage{longtable}
\usepackage{amsthm}
\usepackage{amssymb }
\usepackage{amsmath}
\usepackage{bm}
\usepackage{mathrsfs}
\usepackage{amsfonts}
\usepackage{longtable}
\usepackage{multirow}
\usepackage{algorithm, algpseudocode}
\usepackage{cite}
\usepackage{stfloats}

\title{A Particle-Filtering Based Approach for Distributed Fault Diagnosis of Large-Scale Interconnected Nonlinear Systems}
\author{
\authorblockN{Elaheh Noursadeghi, Ioannis Raptis}\\
\authorblockA{Mechanical Engineering Department, University of Massachusetts Lowell\\
  Email: $\{$Elaheh\_Noursadeghi, Ioannis\_Raptis$\}$@uml.edu}
}

\begin{document}

\maketitle

\begin{abstract}
This paper deals with the problem of designing a distributed fault detection and isolation algorithm for nonlinear large-scale systems that are subjected to multiple fault modes. To solve this problem, a network of detection nodes is deployed to monitor the monolithic system. Each node consists of an estimator with partial observation of the system's state. The local estimator executes a distributed variation of the particle filtering algorithm; that process the local sensor measurements and the fault progression model of the system. In addition, each node communicates with its neighbors by sharing pre-processed information. The communication topology is defined using graph theoretic tools. The information fusion between the neighboring nodes is performed by a distributed average consensus algorithm to ensure the agreement on the value of the local estimates. The simulation results demonstrate the efficiency of the proposed approach. 
\end{abstract}


\section{Introduction}

The majority of the systems that are frequently encountered in all aspects of our everyday life; describing technological, environmental, financial and social processes are characterized by considerable size and significant complexity. In the technical world, contemporary industrial processes are composed by a large number of spatially distributed feedback modules with heterogeneous sensors, actuators and controllers that exchange information over a band-limited communication network that is embedded to the system. These systems are characterized as large-scale networked control systems and can be found in many real-life applications such as: telecommunication networks, water distribution systems, traffic networks, power systems, multi-vehicle formations to name a few. 

Every system is susceptible to faults that may lead to catastrophic failures. Complex processes are significantly more vulnerable to faults, since a malfunction in a single component may have a major effect to the entire system. There is a growing need for reliable real-time monitoring and supervision especially in the case of safety-critical systems. The broad objective is to design fault tolerant systems that maintain their operation even in the occurrence of faults. A timely diagnosis of a fault mode may improve the system's availability and maintainability by avoiding down-times, breakdowns and catastrophic failures.

Fault diagnosis has received considerable attention since the 1970s. Most of the existing techniques involve a centralized architecture \cite{Kabore2001,Zhang2001,Jiang2004,Xu2004,Chen2007,Tang2007}, where a single diagnostic module is responsible of receiving, and processing all the information measured by the sensors. This architecture is appropriate for small and centralized systems, however, it is ill-suited for large-scale systems with spatially distributed components. Every monitoring system has certain limitations in terms of computational power and communication bandwidth. When the dimensionality and complexity of the system increases, it is likely that these limitations will not be satisfied by a centralized configuration. 

It is of great importance to express the formulation of fault diagnosis in a non-centralized way, making it applicable to real-life large-scale systems \cite{Zhang2012,Ferrari2012,Boem2013a,Shames2011,Daigle2007,davoodi2014distributed}. Existing methodologies interchange different types of data between their detecting nodes such as: state estimates \cite{yan2008robust,Zhang2012,Daigle2007}, raw measurements of the interconnected states \cite{Ferrari2012,Boem2013a,Shames2011},
or fault signatures \cite{Daigle2007}. The first distributed fault diagnosis approach using particle filtering was introduced in \cite{cheng2005distributed,Cheng2009}. In general, the majority of the existing distributed techniques are designed for discrete-event systems \cite{Baroni1999,rish2005adaptive,le2006graphical}. 

In this work, we present a full-order Distributed Particle Filtering Fault Diagnosis (DPFFD) algorithm for large-scale nonlinear/non-Gaussian systems. The design objective is to develop a network of interconnected Detector Nodes (DNs) to monitor a stochastic nonlinear process that is subject to a set of active fault modes. The DNs should detect distributively the occurrence of a fault by computing the probability of failure of each mode. Each DN has access to a partial and noisy measurement of the system's state and to processed statistical information from its neighboring nodes. It consists of an embedded processing unit that computes a local PF algorithm and a consensus filter that fuses the processed information of neighboring DNs such that the entire detecting system reaches an agreement of its estimates. Graph-based abstractions will be used to represent the active communication channels between the nodes. The output of the detection network is a filtered estimate of the processӳ state and a probability of failure for each fault mode.  

The paper is organized as follows: A brief description of the centralized
particle filtering algorithm is presented in Section \ref{sec:Centralized-Particle-Filtering}. A
centralized particle filtering fault diagnosis approach is analytically introduced in Section \ref{sec:Centralized-Particle-Filtering-1}. In Section \ref{sec:Distributed-Particle-Filtering} the DPFFD algorithm is derived. The performance of the proposed methodology is evaluated in Section \ref{sec:Numerical-Results} via numerical simulations. Finally, concluding remarks are given in Section \ref{sec:Conclusion}.

\section{Particle Filtering} \label{sec:Centralized-Particle-Filtering}

The filtering problem is formulated based on the discrete time state-space
approach. The main concept is to estimate the system's state by using
a sequence of noisy measurements. It is assumed that the system's
measurements are available at every discrete time instant. The state
estimation should take place recursively as the measurements are received.
Consider a time-dependent, state vector $x(k)\in\mathbb{R}^{n_{x}}$,
where $k\in\mathbb{Z}^{+}$ is the time index. The state-transition
model of the state $x$ is defined according to:
\begin{equation}
\begin{aligned}x(k+1) & =f(x(k))+v(k)\end{aligned}
\label{eq:state transition model}
\end{equation}
where $f:\mathbb{R}^{n_{x}}\times\mathbb{R}^{n_{v}}\rightarrow\mathbb{R}^{n_{x}}$
is a known, nonlinear function, and $v\in\mathbb{R}^{n_{v}}$ stands
for the system's noise. At time step $k$, the observation equation
is expressed by:
\begin{equation}
\begin{aligned}z(k) & =h(x(k))+\omega(k)\end{aligned}
\label{eq:local measurement model}
\end{equation}
where $z\in\mathbb{R}^{n_{z}}$ represents the measurement vector,
$h:\mathbb{R}^{n_{x}}\times\mathbb{R}^{n_{\omega}}\rightarrow\mathbb{R}^{n_{z}}$
is a known nonlinear function, and $\omega\in\mathbb{R}^{n_{\omega}}$
stands for the measurement noise. It is assumed that both the system
noise $v$ and the measurement noise $\omega$ are white and independent. 

In real life applications, most systems are nonlinear and driven by
non-Gaussian noise. As a result, optimal filters such as the Kalman
Filter (KF) and its derivatives become ill-suited. Particle Filters
(PFs) is an alternative to the Kalman approach, that is suitable to
nonlinear problems. The PF is a sequential Bayesian estimator that
utilizes Monte Carlo (MC) simulations. This approach enables us to
represent the non-Gaussian posterior probability density function
(pdf) $p(x(k)|z(k))$ by a set of $N_{s}$ randomly drawn particles
$x^{i}(k)$ and their corresponding weights $w^{i}(k)$. Based on
this representation, the posterior can have the following discrete
approximation:
\begin{equation}
\hat{p}(x(k)|z(k))\approx\sum_{i=1}^{N_{s}}w^{i}(k)\delta(x(k)-x^{i}(k))\label{eq:posterior-aprox}
\end{equation}
where $\delta(\cdot)$ denotes to the multivariate delta function.
A way to mitigate the inability to directly sample from a posterior
distribution is to apply the methodology of Sequential Importance
Sampling (SIS). This technique allows the computation of a distribution
using random samples that are drawn from another. The particles $x^{i}\left(k\right)$
are drawn from an importance density function (proposal pdf) $q\left(x\left(k\right)|x\left(k-1\right),z\left(k\right)\right)$.
Their corresponding weights $w^{i}\left(k\right)$, are sequentially
calculated involving the likelihood function according to:
\begin{equation}
w^{i}(k)\propto w^{i}(k-1)\frac{p\big(z(k)|x^{i}(k)\big)p\big(x^{i}(k)|x^{i}(k-1)\big)}{q\big(x^{i}(k)|x^{i}(k-1),z(k)\big)}\label{eq:weights}
\end{equation}
Filtering via SIS takes place by recursively updating the importance
weighs $w^{i}$ and the particles $x^{i}$ as new measurements $z$
become available. The most basic form of the PF algorithm is the Sequential
Importance Resampling Filter (SIR). The most popular SIR algorithm
is the bootstrap filter that uses the state-transition density $p\left(x\left(k\right)|x\left(k-1\right)\right)$
as the importance pdf. In this case, the update weight equation simplifies
to $w^{i}\left(k\right)\propto w^{i}\left(k-1\right)p\left(z\left(k\right)|x^{i}\left(k-1\right)\right)$. 

\begin{figure}
\centering{}\includegraphics[width=1\columnwidth]{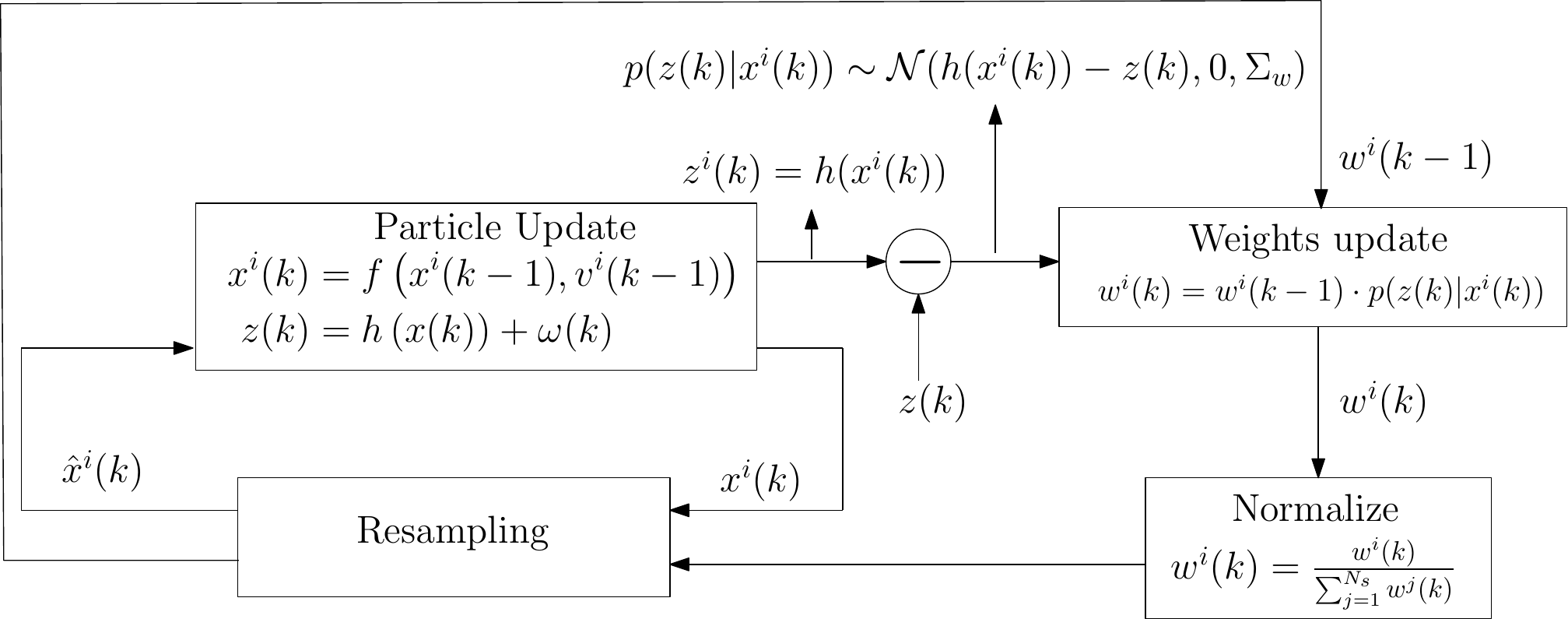}\caption{Block diagram of the bootstrap filter. \label{fig:Schematic-of-PF} }
\end{figure}

The particles will degenerate rapidly since the transition prior is
not conditioned on the measurement data. The transition prior $p\left(x\left(k\right)|x\left(k-1\right)\right)$
is a broader distribution than the likelihood $p(z\left(k\right)|x\left(k\right))$,
thus, only a few particles are assigned with a high weight. The performance
of the algorithm will deteriorate rapidly and especially when the
measurement noise is small. In order to make the SIS simulation-based
techniques viable, the \emph{resampling }method of particles is adopted.
The fundamental idea behind resampling is to preserve particles with
large weights while discarding those with small weights. More about
resampling techniques can be found in \cite{ristic2004beyond}. The
schematic of the bootstrap filter is depicted in Figure \ref{fig:Schematic-of-PF}.

\section{Centralized Particle Filtering Fault Detection and Isolation}\label{sec:Centralized-Particle-Filtering-1}

This Section presents a Centralized Particle Filtering Fault Detection
(CPFFD) algorithm for nonlinear and non-Gaussian stochastic nonlinear
processes. In this work we extent the methodology reported in \cite{orchard2007particle}
from simple one dimensional fault-growth models to state-space representations
of generic nonlinear systems. The main objective of the CPFFD methodology
is to provide an estimate of the system's state by a noisy measurements
sequence (filtering), and generate a statistical characterization
of the fault mode that can trigger fault alarms. Consider the uncertain,
nonlinear and discrete-time dynamic system $S$, described by the
following state-space model:
\begin{equation}
S:\begin{aligned}x\left(k+1\right) & =f(x(k),u(k))+\sum_{j=1}^{M}\beta(k-k_{0}^{j})\cdot g^{j}(x(k),u(k))\\
 & \qquad\qquad\qquad\quad+v(k)\\
z\left(k\right) & =h\left(x\left(k\right)\right)+\omega\left(k\right)
\end{aligned}
\label{eq:state space model}
\end{equation}
\noindent where the terms $x\left(k\right)\in\mathbb{R}^{n_{x}}$,
$u\left(k\right)\in\mathbb{R}^{n_{u}}$, and $z\left(k\right)\in\mathbb{R}^{n_{z}}$
refer to the state, input and measurement vector, respectively; $f:\mathbb{R}^{n_{x}}\times\mathbb{R}^{n_{u}}\rightarrow\mathbb{R}^{n_{x}}$,
and $h:\mathbb{R}^{n_{x}}\rightarrow\mathbb{R}^{n_{z}}$ denote the
known nonlinear functions of healthy and measurement dynamics of the
system, while $v\left(k\right)$ and $\omega\left(k\right)$ stand
for the system and measurement noise with appropriate dimensions,
respectively. 

The term $\{g^{j}\left(x\left(k\right),u\left(k\right)\right)\}_{j=1}^{M}$
refer to $M$ potential fault modes where the nonlinear function $g^{j}:\mathbb{R}^{n_{x}}\times\mathbb{R}^{n_{u}}\rightarrow\mathbb{R}^{n_{x}}$
represents the dynamic of fault mode $j$. The term $\beta(k-k_{0})$
is a scalar function representing the time profi{}le of the fault
occurrence that takes place at some unknown time $k_{0}$. Here, we
can consider both types of faults: \emph{abrupt} (step-like) or \emph{incipient}
(exponential-like) faults, described by:
\begin{equation}
\begin{aligned}\beta(k-k_{0}) & =\begin{cases}
0 & \quad k<k_{0}\\
\underbrace{1}_{\text{abrupt}}\text{\;\ or}\;\underbrace{1-c^{-\left(k-k_{0}\right)},\: c>1}_{\text{incipient}} & \quad k\ge k_{0}
\end{cases}\end{aligned}
\label{eq:scalar function}
\end{equation}
The algorithm incorporates the process model of (\ref{eq:state space model}),
as well as a binary state vector, to detect changes in the process
dynamics expressed by the terms $\beta(k-k_{0})$. To this end, a
vector of binary states $b^{j}\left(k\right)=[\begin{array}{cc}
b_{1}^{j}\left(k\right) & b_{2}^{j}\left(k\right)\end{array}]^{T}$ with $b_{1},b_{2}\in\left[0\:1\right]$ and $j=1,\ldots,M$, is used
to signal the occurrence of each fault mode. More specifically, $b_{1}^{j}\left(k\right)=1$
indicates that the system is operating normally, while $b_{2}^{j}\left(k\right)=1$
denotes the presence of fault mode $j$ in the system. This technique
enables us to calculate probabilistic measures related to the existence
of a fault in the system. 

To determine the operating condition of the system (normal or faulty
operating condition) and make a decision based on occurrence of faults,
a particle filtering approach will be employed for a statistical characterization
of both the binary and continuous-valued states, as new measurements
are received. The binary states update sequence is augmented to the
state-transition model $S$ given in (\ref{eq:evolution of continuous state}).
This way, the state-transition model of system $S$ is converted to
a \emph{failure sensitive filter} that reflects the presence of changes
to the original healthy dynamics. Hence, the state vector that is
used by the PF algorithm is $\mathcal{X}^T(k) = [\begin{array}{cccc}
 (x^{c}(k)) ^{T} & (b^{1} (k)) ^{T} & \cdots & (b ^{M}(k)) ^{T} \end{array}] \in \mathbb{R} ^{n_{x}+2 \cdot M}$, where $x^{c} (k) = x(k)$. The state-space model of the system that is implemented in the PF algorithm can be
expressed in terms of the continuous states vector, and the binary
states vector. Therefore, the state-transition model of these two
state vectors and the measurement equation can be written as:
\begin{equation}
\begin{aligned}x^{c}\left(k+1\right) & =f\left(x^{c}\left(k\right),u\left(k\right)\right)+\sum_{j=1}^{M}g^{j}\left(x^{c}\left(k\right),u\left(k\right)\right)\cdot b_{2}^{j}\left(k\right)\\
 & \qquad+\tilde{v}\left(k\right)\\
b^{j}\left(k+1\right) & =\Phi\left(b^{j}\left(k\right)+n^{j}\left(k\right)\right)\text{ }\qquad\qquad j=1,\ldots M\\
z\left(k+1\right) & =h\left(x^{c}\left(k\right)\right)+\tilde{\omega}\left(k\right)
\end{aligned}
\label{eq:evolution of continuous state}
\end{equation}
\noindent where $\tilde{v}\left(k\right)$, and $\tilde{\omega}\left(k\right)$
are the approximations of non-Gaussian process and measurement noise
with appropriate dimensions, respectively. These noise sequences should
be as close as possible to the actual ones ($v\left(k\right)$ and
$\omega\left(k\right)$). The evolution of the binary states $\Phi:\mathbb{R}^{2}\rightarrow\{[0\;1]^{T},[1\;0]^{T}\}$,
is a nonlinear function driven by the identically independent distributed
(i.i.d) uniform white noise $n^{j}\left(k\right)$. The function $\Phi\left(\cdot\right)$
is defined such that the previous state $b^{j}\left(k\right)$ is
randomly excited at each time step by $n\left(k\right)$. This random
vector of $\mathbb{R}^{2}$ is assigned to ones of the binary states
(healthy/faulty) based on the distance metric of the perturbed vector
$b^{j}\left(k\right)+n\left(k\right)$ to the coordinates $[0\;1]^{T}$
and $[1\;0]^{T}$. With this technique, when a fault occurs, the weights
will gradually converge the binary state $b_{2}^{j}\left(k\right)$
to one ($b_{2}^{j}\left(k\right)\rightarrow1$). This is due to the
fact that the likelihood of the measurements will diminish the weights
of particles associated with the healthy condition. A choice of $\Phi\left(\cdot\right)$
that has been successfully used in \cite{orchard2007particle,Orchard2009,raptis2011adaptive,raptisparticle}
is: 
\begin{equation}
\begin{aligned}\Phi(x) & =\begin{cases}
e_{1} & \text{if }\parallel x-e_{1}\parallel\le\parallel x-e_{2}\parallel\\
e_{2} & \text{else}
\end{cases}\end{aligned}
\label{eq:function of Boolean states}
\end{equation}
\noindent where $e_{1}=\left[\begin{array}{cc}
1 & 0\end{array}\right]^{T}$ and $e_{2}=\left[\begin{array}{cc}
0 & 1\end{array}\right]^{T}$. The state model of the CPFFD algorithm can be written in a more
compact form as:
\begin{align}
\mathcal{X}\left(k+1\right) & =\mathcal{F}\left(\mathcal{X}\left(k\right),u\left(k\right),\mathcal{V}\left(k\right)\right)\label{eq:compact-form}\\
\mathcal{Z}\left(k\right) & =\mathcal{H}\left(\mathcal{X}\left(k\right)\right)+\tilde{\omega}\left(k\right)\nonumber 
\end{align}
\noindent where $\mathcal{V}=\left[v^{T}\;\big(n^{1}\big)^{T}\ldots\big(n^{M}\big)^{T}\right]^{T}$,
$\mathcal{Z}=z$, and $\mathcal{F}(\cdot)$, $\mathcal{H}(\cdot)$
are nonlinear functions of appropriate dimensions and structure calculating
based on (\ref{eq:evolution of continuous state}). The above definition
will be used to ease the notation in the subsequent parts of this
paper. The outputs of the CPFFD module are the \emph{probabilities
of each failure mode}. These are the expectations of the binary states
$\hat{b}_{2}^{j}\left(k\right)=E\left[b_{2}^{j}\left(k\right)|z\left(k\right)\right]$.
This value is used to trigger alarm indicators if its value exceeds
a certain threshold $\alpha\in\left(0\:1\right)$ that marks the probability
of detection (i.e. $\hat{b}_{2}^{j}\left(k\right)<\alpha$ indicates
normal operation). With this layout two or more different co-existing
fault modes can be simultaneously detected.

\section{Distributed Particle Filtering Fault Diagnosis }\label{sec:Distributed-Particle-Filtering}

The CPFFD algorithm described in the previous Section is not scalable
or robust for complex large-scale dynamical systems that employ scattered
measurement sensors over large geographical regions. 



\begin{figure}
\centering{}\includegraphics[width=1\columnwidth]{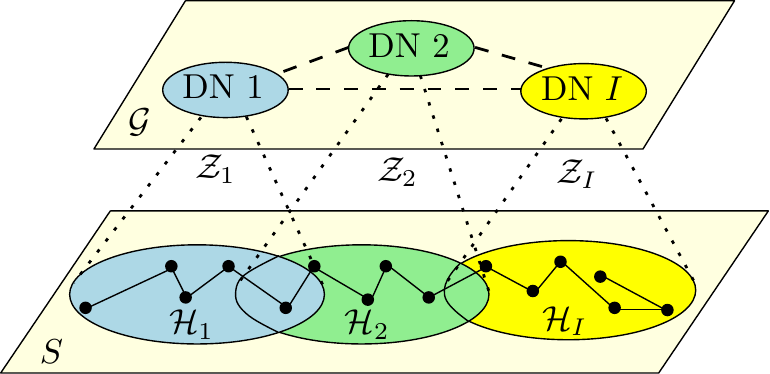}\caption{Block diagram of DPFFD approach \label{fig:Proposed-distributed-fault}}
\end{figure}

Here, we propose a Distributed Particle Filtering Fault Diagnosis
(DPFFD) algorithm for large-scale nonlinear/non-Gaussian systems.
The main objective is to develop a network of interconnected Detector
Nodes (DNs) to monitor the full-order system that is subject to a
set of active fault modes. Figure \ref{fig:Proposed-distributed-fault} indicates the block
diagram of the proposed distributed approach. Similar to Section \ref{sec:Centralized-Particle-Filtering}, we assume the presence of $M$
potential fault modes in system $S$ (\ref{eq:state space model}).
A network of $N$ DNs is deployed to monitor the full-order system.
The local measurement equation of each of which is expressed by:
\begin{equation}
\begin{aligned}z_{I}(k) & =h_{I}(x(k))+\omega_{I}(k)\qquad I=1,2,...,N\end{aligned}
\label{eq:local measurement model-1}
\end{equation}
where $z_{I}(k)\in\mathbb{R}^{n_{z_{I}}}$ refers to the observation
vector of DN $I$; the nonlinear function $h_{I}:\mathbb{R}^{n_{x}}\rightarrow\mathbb{R}^{n_{z_{I}}}$
indicates the local measurement dynamics, and $\omega_{I}$ stands
for its corresponding measurement noise.

Analogously to (\ref{eq:compact-form}), a compact formulation of
the state-transition model that is used by the local PF module at each
DN can be written as: 
\begin{equation}
\begin{aligned}\mathcal{X}(k+1) & =\mathcal{F}(\mathcal{X}(k),u(k),\mathcal{V}(k))\\
\mathcal{Z}_{I}(k) & =\mathcal{H}_{I}(\mathcal{X}(k))+\hat{\omega}_{I}(k)
\end{aligned}
\label{eq:compact form LD}
\end{equation}
\noindent where $\mathcal{Z}_{I}=z_{I}$; $\mathcal{F}(\cdot)$ is
calculated according to (\ref{eq:evolution of continuous state}),
and $\mathcal{H}_{I}(\cdot)$ is a nonlinear function that represents
the local measurement dynamic (\ref{eq:local measurement model-1}).
It is reminded that $\mathcal{X}$ includes both the continuous system's
states vector $x^{c} = x$ as well as the binary states vector
$b ^{j}\left(k\right)=[\begin{array}{cc}
b_{1} ^{j}\left(k\right) & b_{2} ^{j}\left(k\right)\end{array}]^{T}\: j=1,\ldots,M$. 

As mentioned, the goal of the DPFFD algorithm is to estimate the full
system's state $\mathcal{X}(k)$ in a sequential manner and diagnose
distributively the occurrence of the fault modes based on the local
measurement vector of each DN. 

Graph-based abstractions will be used to represent the active communication channels between the nodes. The communication network will be described by the graph $\mathcal{G}$, defined as the pair $\mathcal{G=}(\mathcal{V},\mathcal{E})$, where $\mathcal{V}=\{v_{1},\ldots,v_{N}\}$ is the vertex set or DNs set, and $\mathcal{E}=\{\{v_{i},v_{j}\}\in\mathcal{V}\times\mathcal{V}\}$ represents the set of edges of $\mathcal{G}$. Each element of $\mathcal{E}$ represents an undirected communication link between two DNs. The neighborhood $\mathcal{N}_{i}\subseteq\mathcal{V}$ of the vertex $v_{i}$ is defined as the set of all vertices that are adjacent to $v_{i}$, $\{v_{j}\in\mathcal{V}|\{v_{i},v_{j}\}\in\mathcal{E}\}$. If $v_{j}\in\mathcal{N}_{i}$, it follows that $v_{i}\in\mathcal{N}_{j}$, since an undirected edge exists between them.

The algorithm is performed in several steps at each DN in parallel
with other neighboring nodes. Each DN executes a local bootstrap PF
(\emph{particle update}). In the first step, $N_{s}$ particles are
drawn according to the state transition equation of (\ref{eq:compact form LD}). 

The next step is the \emph{weight update step}. In this step, each
node uses its local observation and communicates with its neighbors
to update its particle weights based on $p\left(\mathcal{Z}(k)|\mathcal{X}^{i}\left(k\right)\right)$.
The main problem here is that each DN does not have access to global
observation vector (it is defined as the union of local observation
vectors, $\mathcal{Z}(k)=\bigcup_{I=1}^{N}\mathcal{Z}_{I}(k)$.
To calculate the value of $p\left(\mathcal{Z}(k)|\mathcal{X}^{i}\left(k\right)\right)$
in a distributed manner at each DN, each node requires to calculate
its local likelihood function $p\left(\mathcal{Z}_{I}(k)|\mathcal{X}^{i}\left(k\right)\right)$
based on its observation and communicates with its neighboring nodes.

\begin{figure}
\centering
\includegraphics[scale=0.5]{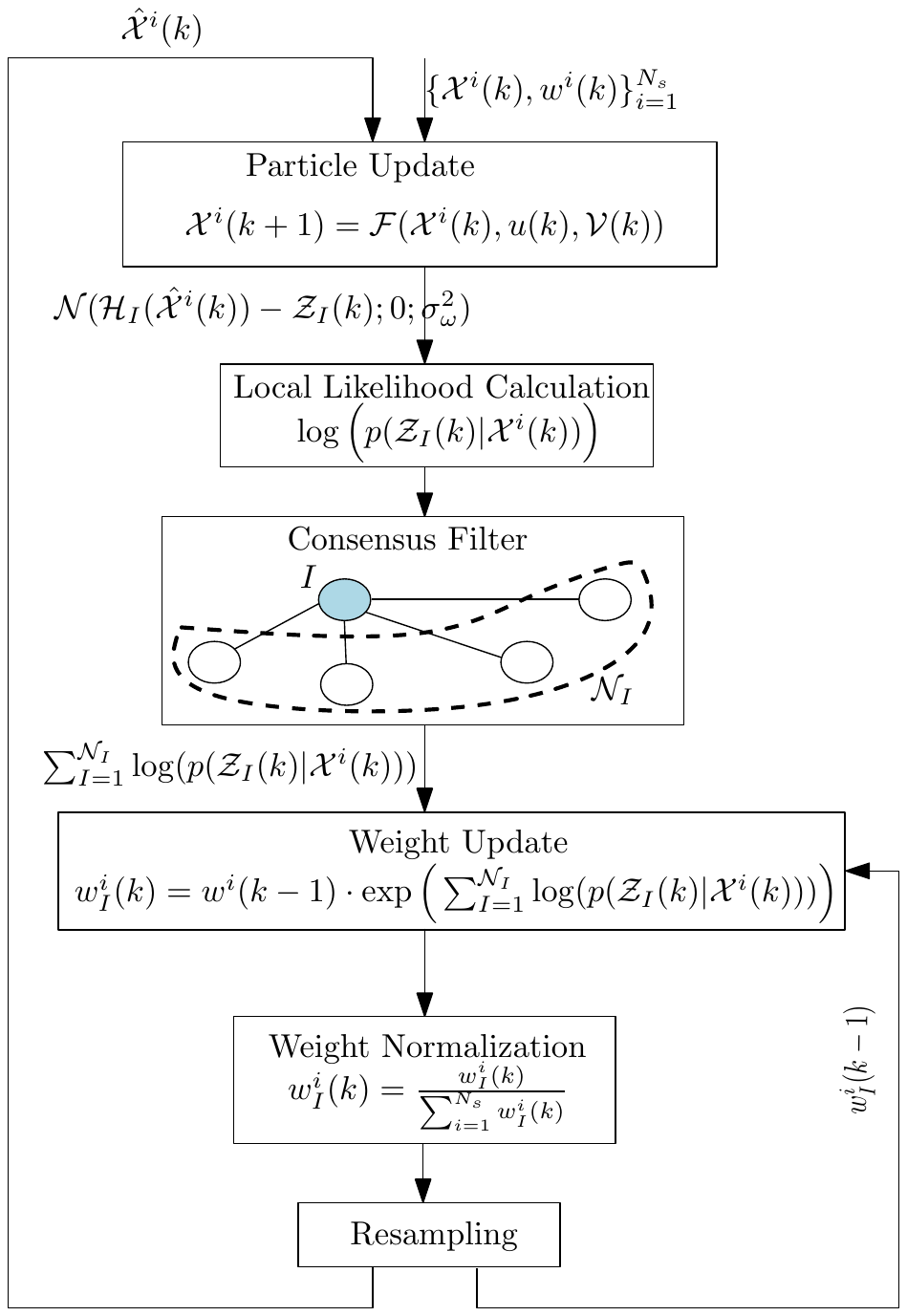}
\caption{Schematic of the DPFFD algorithm's execution flow in DN $I$. \label{fig:Block-diagram-of}}
\end{figure}

To calculate the value of $p\left(\mathcal{Z}(k)|\mathcal{X}^{i}\left(k\right)\right)$
in a distributed manner, we assume that the local measurement noise
sequences of the DNs are independent. Therefore, the global likelihood
function can be factorized as the product of the local likelihood
distributions \cite{hlinka2013distributed,coates2004distributed,sheng2005distributed}
as: 
\begin{equation}
\begin{aligned}p(\mathcal{Z}(k)|\mathcal{X}(k)) & =\prod_{I=1}^{N}p(\mathcal{Z}_{I}(k)|\mathcal{X}(k))\end{aligned}
\label{eq:global likelihood-1}
\end{equation}
The above equation requires the same set of particles $\{\mathcal{X}_{I}^{i}(k)\}_{i=1}^{N_{s}}=\{\mathcal{X}^{i}(k)\}_{i=1}^{N_{s}}\; I=1,\ldots,N$
is sampled at each iteration, hence, the synchronization of the local
random number generators of DNs is necessary. The synchronous operation
of the DNs can be achieved by use of random number generators with
identical seeds that are initialized at the same points. Therefore,
each DN calculates its local likelihood function based on its observation
vector. Following (\ref{eq:weights}) and the factorization of the
global likelihood distribution given in (\ref{eq:global likelihood-1}),
the particle weights are calculated as follows:
\begin{equation}
\begin{aligned}w_{I}^{i}(k) & =w^{i}(k-1)\cdot\left(\prod^{J\in\{I,\mathcal{N}_{I}\}}p(\mathcal{Z}_{J}(k)|\mathcal{X}^{i}(k))\right)\end{aligned}
\label{eq:weight update-1}
\end{equation}
where {\small{}$\prod^{J\in\{I,\mathcal{N}_{I}\}}p(\mathcal{Z}_{J}(k)|\mathcal{X}^{i}(k))$}
indicates the product of the local likelihood functions of node $I$
and its neighboring nodes represented by $\mathcal{N}_{J}$. Since,
it is often easier to work with the logarithm of a probability rather
than the probability itself, we take the natural logarithm from (\ref{eq:weight update-1}).
Therefore, applying the natural logarithm \cite{hlinka2013distributed}
yields to:
\begin{equation}
\begin{aligned}\log\left(\prod_{I=1}^{N}p(\mathcal{Z}_{I}(k)|\mathcal{X}^{i}(k))\right) & =\sum_{I=1}^{N}\log p(\mathcal{Z}_{I}(k)|\mathcal{X}^{i}(k))\end{aligned}
\label{eq:logarithm}
\end{equation}
The above sum can be calculated in a distributed way by means of distributed
average consensus algorithm \cite{xiao2004fast,xiao2005scheme} and
communication graph topology. The consensus filter is executed iteratively
in every time step of the DPFFD algorithm. The input of the consensus
filter at each DN is the logarithm of the local likelihood function,
$\log p(\mathcal{Z}_{J}(k)|\mathcal{X}^{i}(k))$. In the consensus
algorithm, each DN computes $\sigma_{I}=\sum^{J\in\{I,\mathcal{N}_{I}\}}\Big(\log p(\mathcal{Z}_{J}(k)|\mathcal{X}^{i}(k))\Big)$
iteratively by communicating with its neighbors across the pre-defined
graph topology and use of the weighted average as:
\begin{equation}
\begin{aligned}\sigma_{I}(t+1) & =\mathcal{W}_{II}\sigma_{I}(t)+\sum_{J\in\mathcal{N}_{I}}\mathcal{W}_{IJ}\sigma_{J}(t)\end{aligned}
\quad I=1,\ldots,N\label{eq:consensus algorithm}
\end{equation}
\noindent where $\mathcal{W}_{II}$ and $\mathcal{W}_{IJ}$ are fixed
weights.We employ the Metropolis weights to calculate the value of
$\mathcal{W}_{IJ}$ \cite{xiao2005scheme,xiao2004fast}. 

Note that the time index of the consensus filter is denoted
by $t$ indicating that the algorithm is iterated at every time step
$k$ of the local PF algorithm. The algorithm is executed until a
certain convergence error is achieved or a pre-determined number of
iterations is reached. In fact, the consensus filter plays the role
of observation fusion center. Its output relies on two factors: i)
type of the consensus algorithm; and ii) the communication graph topology
between DNs. Finally the particle weights are updated according to:
\begin{equation}
\begin{aligned}w_{I}^{i}(k) & =w^{i}(k-1)\cdot\exp\left(\sum^{J\in\{I,\mathcal{N}_{I}\}}\log p(\mathcal{Z}_{J}(k)|\mathcal{X}^{i}(k))\right)\end{aligned}
\label{eq:global weight update}
\end{equation}
The schematic of the local PF algorithm performed in one of the DNs
is depicted in Figure \ref{fig:Block-diagram-of}. Other steps include:
weight normalization, resampling, and MMSE state estimate performed
in the same way as explained in Section \ref{sec:Centralized-Particle-Filtering}.

\begin{figure}
\centering{}\includegraphics[width=1\columnwidth]{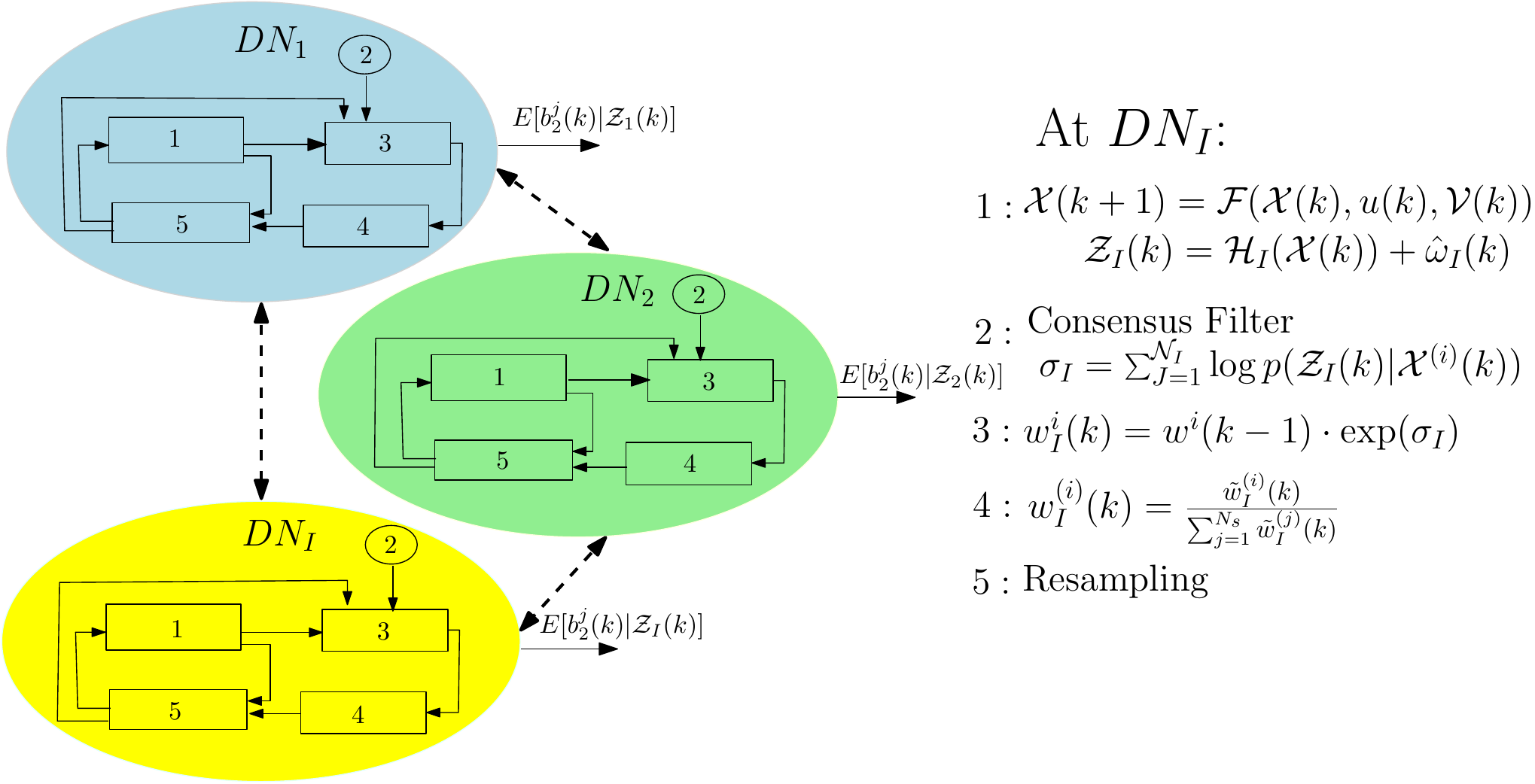}\caption{Block diagram of the DPFFD algorithm. \label{fig:Schematic-of-distributed}}
\end{figure}

The output of each DN is the probabilities of the failure modes $\hat{b_{2}}^{j}\left(k\right)=E\left[b_{2}^{j}\left(k\right)|\mathcal{Z}_{I}\left(k\right)\right]$.
It is important to note that due to the consensus filter embedded
at each DN, we have $E\left[b_{2}^{j}\left(k\right)|\mathcal{Z}_{I}\left(k\right)\right]=E\left[b_{2}^{j}\left(k\right)|\mathcal{Z}\left(k\right)\right]$.
This equality states that the probability of each failure mode has
the same value in every DN due to the execution of the distributed
agreement protocol. Similar to the centralized case, this value is
used to trigger alarm indicators if it exceeds a certain threshold
$\alpha\in\left(0\:1\right)$ that marks the probability of detection.
The block diagram of the DPFFD algorithm is depicted in Figure \ref{fig:Proposed-distributed-fault}.
During the execution of the diagnostic routine, each DN computes its
local likelihood function based on a reduced-order measurement of
the system's state. Then, each DN broadcasts its local processed data
to the consensus filter that is used to compute the particle weights.
An illustration of the full-order DPFFD algorithm is depicted in Figure
\ref{fig:Schematic-of-distributed}. The pseudo code is provided in Table \ref{tab:Fault-identification-algorithm}.

\begin{table}
\caption{DPFFD algorithm. \label{tab:Fault-identification-algorithm}}
\begin{algorithmic}[1]
\For {$I=1:N$}
\Statex \textbf{Initialization:}
\State $\mathcal{X}^T(0)=[\underbrace{1\,0}_{\text{mode 1}}  ...\underbrace{1\,0}_{\text{mode j}} ... \underbrace{1\,0} _{\text{mode M}}\, x_c ^T(0)]$
\State $\text{At time}\;k\ge 1$
\Statex \textbf{Particle Update:}
\State $\mathcal{X}(k+1)=\mathcal{F}(\mathcal{X}(k),u(k),\mathcal{V}(k))$
\Statex \textbf{Weight Update:}
\State $w_I(k)=w(k-1)\cdot\prod_{I=1} ^N p(\mathcal{Z}_I(k)|\mathcal{X}^i (k))$
\State $\log\left(\prod_{I=1} ^N p(\mathcal{Z}_I(k)|\mathcal{X}^i (k))\right)=$\\
\Statex $\underbrace{\exp(\sum_{I=1} ^N \log p(\mathcal{Z}_I(k)|\mathcal{X}^i (k)))}_{\text{distributed average consensus}}$
\Statex \textbf{Weight Normalization:}
\State $w^i _I(k)=\frac{w^i _I(k)}{\sum_{i=1} ^{N_s} w^i _I(k)}$
\Statex \textbf{Resampling}
\Statex \textbf{Calculation of MMSE state estimate:}
\State $\tilde{\mathcal{X}}(k)=\sum_{i=1} ^{N_s} w^i _I(k)\hat{\mathcal{X}}^i(k) $
\Statex \textbf{Calculation the probability of each failure mode:}
\For {$j=1:M$}
\State $E[b_2 ^j(k)|\mathcal{Z}(k)]=\frac{\sum_{i=1} ^{\text{number of particles with}\;b_2 ^j(k)=1} w^i _I(k)}{\sum_{j=1} ^{N_s} w^j _I(k)}$
\EndFor

\EndFor
\end{algorithmic}
\end{table}

\section{Numerical Results}\label{sec:Numerical-Results}

\begin{figure}
\centering
\includegraphics[scale=0.6]{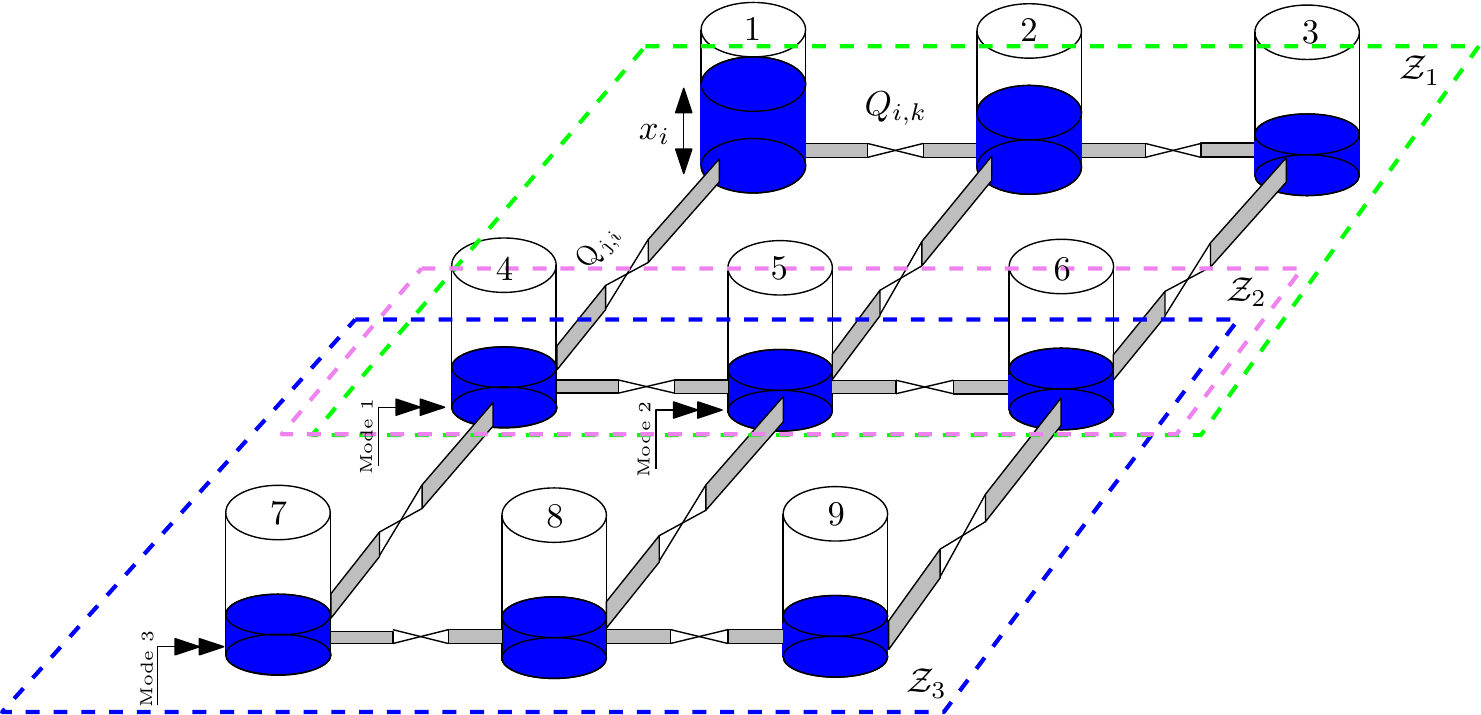}
\caption{Schematic of the nine-tank system monitored by three DNs. The location of the fault nodes is highlighted by the double arrows. \label{fig:Schematic-of-the}}
\end{figure}

The DPFFD algorithm was evaluated via numerical simulations. The system
under investigation is a chemical process consisted of nine identical
interconnected tanks filled with a liquid. The cross section of the
cylindrical tanks and their connection are denoted by $S_{c}$ and
$S_{p}$, respectively. The tanks are arranged in a grid formation
as shown in Figure \ref{fig:Schematic-of-the}. The level of tank
$i$ is denoted by $x_{i}$ and is determined by means of the mass
balance equations as:
\begin{equation}
\begin{aligned}\dot{x}_{i} & =\frac{1}{S_{c}}\Big(\sum Q_{j,i}-\sum Q_{i,k}\Big)+v_{i}\end{aligned}
\label{eq:mass balance equation}
\end{equation}
where $Q_{j,i}$ represents the flow rate from tank $j$ to tank $i$
(inflow rate to tank $i$), and $Q_{i,k}$ refers to the flow rate
from tank $i$ to tank $k$ (outflow rate from tank $i$), and $v_{i}$
stands for the process noise. The flow rate variable, $Q_{i,j}$,
is defined by means of Torricelli's rule as: 
\begin{equation}
\begin{aligned}Q_{i,j} & =\mu_{i}\cdot S_{p}\cdot sign(x_{i}-x_{j})\cdot\sqrt{2g|x_{i}-x_{j}|}\end{aligned}
\label{eq:flow rate definition}
\end{equation}
where the nominal values of the parameters in (\ref{eq:mass balance equation})
and (\ref{eq:flow rate definition}) is given in Table \ref{tab:Model-parameters-of-1}. 

\begin{figure}
\begin{centering}
\includegraphics[width=1\columnwidth]{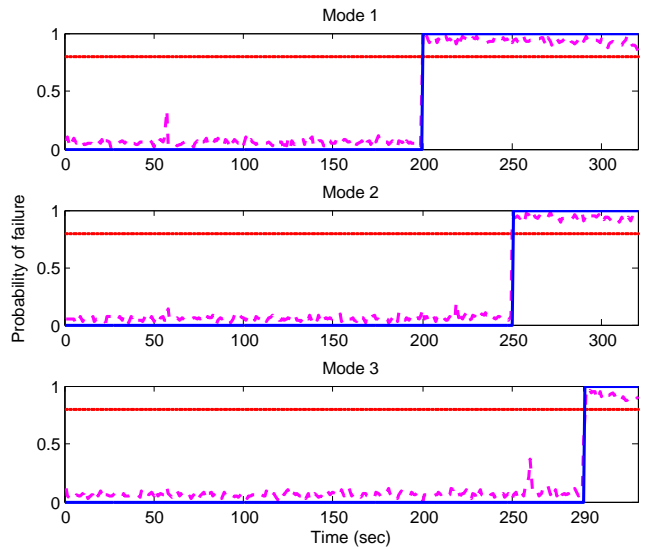}
\par\end{centering}

\caption{Probability of failure for each fault mode calculated by the DNs (dashed magenta line), failure mode occurrence (solid blue line), and threshold level (solid red line).\label{fig:probability-of-failure}}
\end{figure}

\begin{figure}
\centering\includegraphics[width=1\columnwidth]{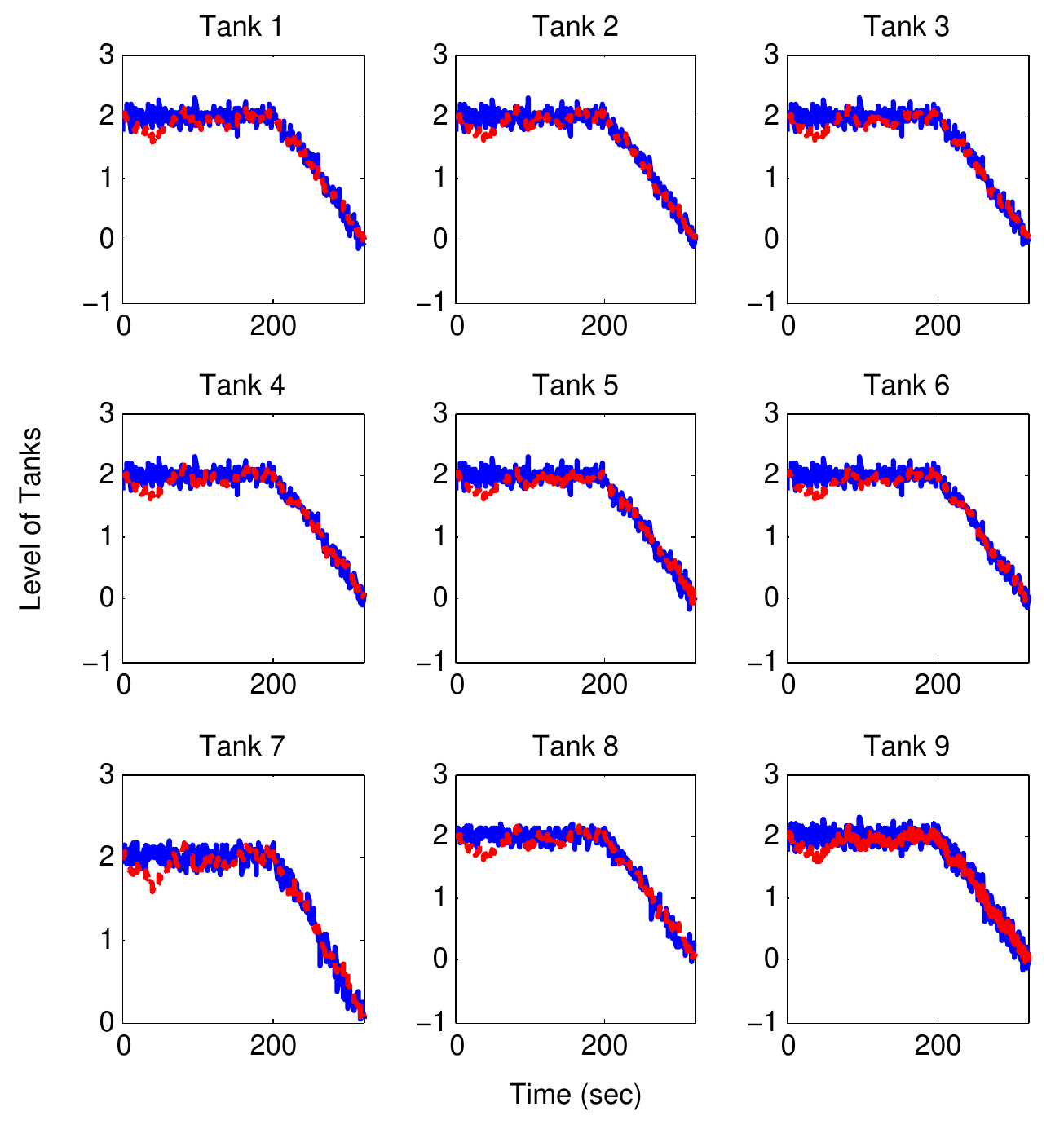}\caption{Comparison of the actual (solid blue line) and estimated value (dashed red line) of the liquid level with respect to time \label{fig:Comparison-of-the}}
\end{figure}

\begin{figure}
\centering\includegraphics[width=1\columnwidth]{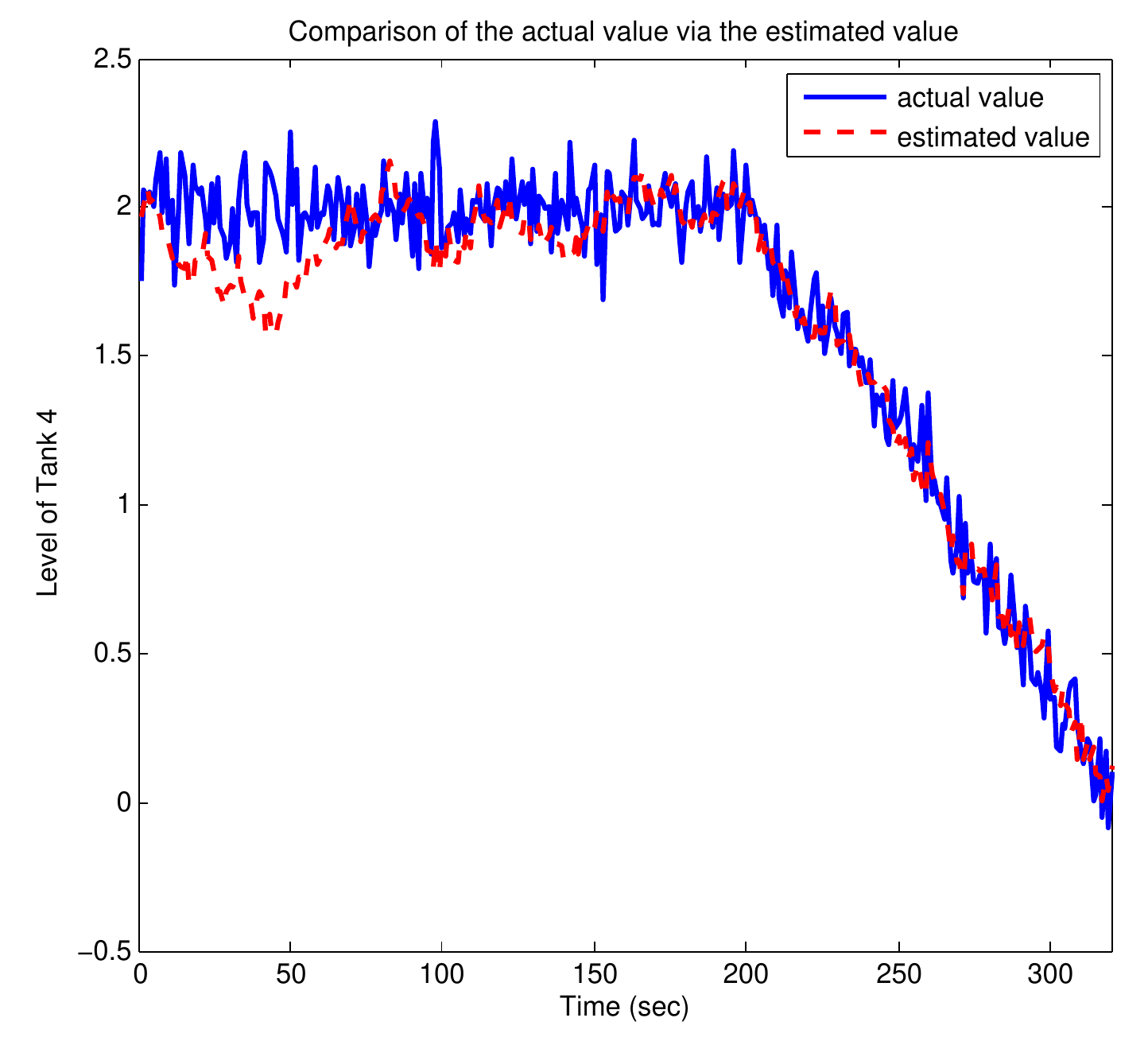}\caption{Comparison of the actual vs the estimated value of the liquid level in tank 4. \label{fig:Comparison-of-the-1}}
\end{figure}

\begin{table}
\caption{Model parameters of the water tank system \label{tab:Model-parameters-of-1}}

\centering{}%
\begin{tabular}{|c|c|c|}
\hline 
Parameter & Meaning & Value\tabularnewline
\hline 
$x_{i}(0)$ & Initial level of tanks & $2\, m$\tabularnewline
$S_{p}$ & Cross section of interacting pipes & $2\times10^{-5}\, m^{2}$\tabularnewline
$S_{c}$ & Tank cross sectional area & $0.0154\, m^{2}$\tabularnewline
$\mu_{i}$ & Flow correction term  & $1$\tabularnewline
$g$ & Gravity constant & $9.81\,\frac{m}{s^{2}}$\tabularnewline
\hline 
\end{tabular}
\end{table}
The three fault modes under consideration are leakages to tanks $4$,
$5$ and $7$ (The fault modes are depicted with double arrows in
Figure \ref{fig:Schematic-of-the}). The leakage model at tank $i$
is described by: 
\begin{equation}
\begin{aligned}g^{i}\left(x_{i}\right) & =\left(\frac{\mu_{i}\cdot S_{p}}{S_{c}}\right)sign(x_{i})\sqrt{2g|x_{i}|}\end{aligned}
\label{eq:fault progression model}
\end{equation}
To apply the DPFFD approach, we discretize the system with a sampling
period of $T_{s}=0.1\: s$. The monolithic system is monitored by
three DNs (dashed lines in Figure \ref{fig:Schematic-of-the}). We
define a \textit{\small{}full-connectivity graph topology} among the
DNs. The state transition and the observation model of the DNs
are expressed as:
\begin{equation}
\begin{aligned}x & =[x_{1}\,\ldots\, x_{9}]^{T}\\
x_{J}(k+1) & =f(x_{J}(k))+v_{J}(k)\quad J=\{1,\ldots,9\}\\
x_{I}(k+1) & =f(x_{I}(k))+\sum_{j=1}^{3}b_{2,I}^{j}\cdot\left(\frac{\mu_{I}\cdot S_{p}}{S_{c}}\right)sign(x_{I})\sqrt{2g|x_{I}|}\\
 & \quad+v_{I}(k)\qquad I=4,5,7\\
z_{1}(k) & =\mathcal{I}_{6}\times[x_{1},\ldots,x_{6}]^{T}+\omega_{1}(k)\\
z_{2}(k) & =\mathcal{I}_{3}\times[x_{4},\ldots,x_{6}]^{T}+\omega_{2}(k)\\
z_{3}(k) & =\mathcal{I}_{6}\times[x_{4},\ldots,x_{9}]^{T}+\omega_{3}(k)
\end{aligned}
\label{eq:observation of each LD}
\end{equation}
where $v_{J}(k)$ is selected as normal distribution, $\mathcal{N}(0,0.05)$,
and $\omega_{I}(k)\; I=1,\ldots,3$ is chosen as multivariate normal
distribution, $\mathcal{N}(0_{n_{z_{I}}},\Sigma)$ with covariance
matrix $\Sigma=a\times\mathcal{I}_{n_{z_{I}}}$ and $\mathcal{I}_{n_{z_{I}}}\in\mathbb{R}^{n_{z_{I}}\times n_{z_{I}}}$
denotes the identity matrix. The positive constant $a=0.2$ is selected
$10\%$ of the nominal value of the tank\textquoteright{}s level. The binary state vector $b_I ^j$ is stimulated by the binary noise in the range $[-0.75\;0.75]$. The probability of each fault mode and the estimation results of the
DPFFD algorithm are depicted in Figures \ref{fig:probability-of-failure}
and \ref{fig:Comparison-of-the}, respectively. The results are the
same for all three DNs. Figure \ref{fig:probability-of-failure} shows
the probability of the different failure modes occurring  in the system
and calculated by the DNs. As it can be seen from this figure, all
three DNs can diagnose the fault modes $1$, $2$, and $3$ at time
instances $T=200,\,250,\,290\, s$ as expected. The comparison of
the real value via the estimated value of the level of the tanks is
illustrated in Figures \ref{fig:Comparison-of-the} and \ref{fig:Comparison-of-the-1}, respectively. Figure \ref{fig:Comparison-of-the-1} indicates the actual and the estimated value corresponding to tank 4. The simulation
results were deemed satisfactory. The DNs accurately and timely identified
all fault modes filtered the system's state. 

\section{Conclusion}\label{sec:Conclusion}
This paper presents a distributed diagnostic module, suitable for large-scale and spatially distributed nonlinear stochastic systems. Instead of a central diagnostic unit, the system is monitored by a set of interconnected detection nodes with local processing and communication capabilities. Graph theoretic tools are deployed to represent the communication topology of the system. Every node executes a consensus protocol that steers the output of the entire network to a common estimate.

 






\end{document}